\documentclass[%
 reprint,
 amsmath,amssymb,
 aps,
prb
]{revtex4-1}

\usepackage{xcolor}
\usepackage{graphicx}
\usepackage{dcolumn}
\usepackage{bm}
\usepackage{braket}

\begin{document}


\title{Supervised-learning approach for recognizing magnetic
skyrmion phases}

\author{I. A. Iakovlev, O. M. Sotnikov, V. V. Mazurenko}
%
\affiliation{
Theoretical Physics and Applied Mathematics Department, Ural Federal University, Mira Street 19, Ekaterinburg 620002, Russia\\
}
\date{\today}

\begin{abstract}
We propose and apply simple machine learning approaches for recognition and classification of complex non-collinear magnetic structures in two-dimensional materials. The first approach is based on the implementation of the single-hidden-layer neural network that only relies on the $z$ projections of the spins. In this setup one needs a limited set of magnetic configurations to distinguish ferromagnetic, skyrmion and spin spiral phases, as well as their different combinations in transitional areas of the phase diagram. The network trained on the configurations for square-lattice Heisenberg model with Dzyaloshinskii-Moriya interaction can classify the magnetic structures obtained from Monte Carlo calculations for triangular lattice and vice versa. The second approach we apply, a minimum distance method performs a fast and cheap classification in cases when a particular configuration is to be assigned to only one magnetic phase. The methods we propose are also easy to use for analysis of the numerous experimental data collected with spin-polarized scanning tunneling microscopy and Lorentz transmission electron microscopy experiments.  
\end{abstract}

\pacs{Valid PACS appear here}
\maketitle


\begin{figure*} 
\center 
\includegraphics[width=180mm]{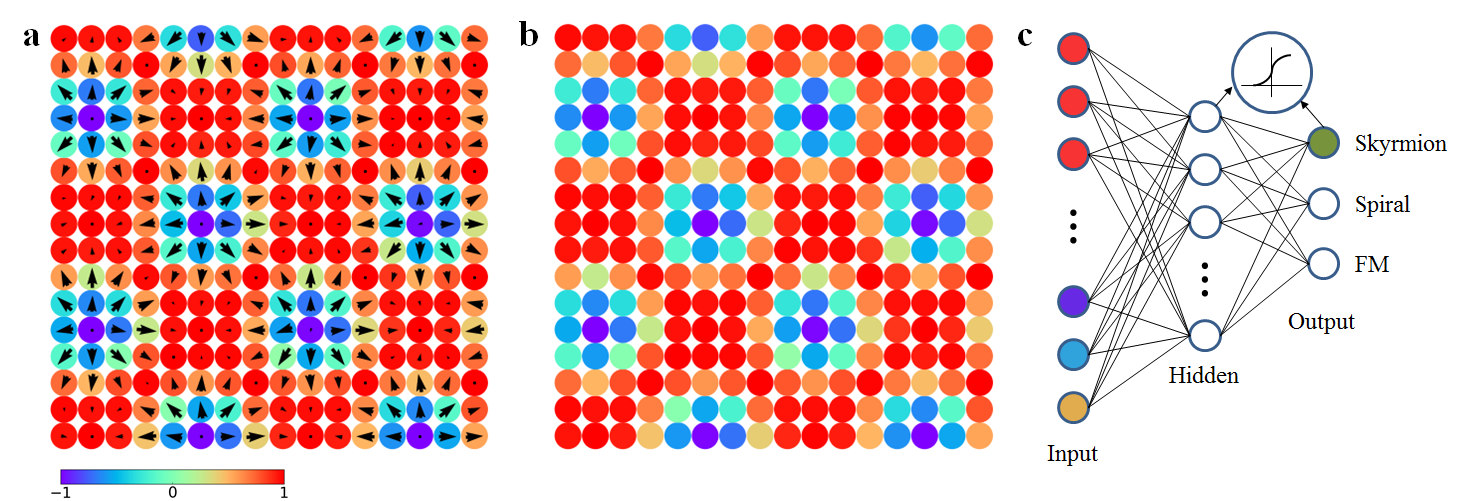} 
\caption{Schematic representation of the machine learning process. (a) The skyrmion magnetic structure as obtained from the classical Monte Carlo simulations for a two-dimensional ferromagnet with Dzyaloshinskii-Moriya interaction at finite temperature and magnetic fields. Black arrows indicate the in-plane $xy$ spin components. (b) The matrix contains the $z$ projection of the spin structure to be classified. (c) Neural network with single hidden layer of sigmoid neurons. The values of the input neurons are equal to $z$ components of the spins of the magnetic configuration.}
\label{neural network} 
\end{figure*}

\section{Introduction}
A fascinating progress in development of neural-network-based approaches in condensed matter theory allows one to advance the methods for studying physical properties of materials. For instance, a neural network representation of the quantum Hamiltonian wave function proposed by Carleo and Troyer \cite{Troyer} has revolutionized the field of simulation of complex many-body systems \cite{Imada, Kato}. Within such an approach it becomes possible to model frustrated systems for which existing Quantum Monte Carlo methods fail due to the sign problem.  Another remarkable example of the innovations in artificial neural network learning is identification of the magnetic phases of the spin Hamiltonians widely used for description of the strongly correlated materials \cite{phase1,phase2,phase3,phase4,phase5,phase6,phase7}. For instance, in the case of the two-dimensional Ising model the ferromagnetic and paramagnetic phases can be successfully recognized with a single-hidden-layer network \cite{Melko1}.  Importantly, topological phases obtained with a more complex XY Hamiltonian \cite{Melko2, Wessel} can be also classified with machine learning, however, in this case one needs to design a deep convolutional network and use the system of filters, which makes such an approach similar to the image recognition \cite{image}.  

Therefore, an important question arises. Is it possible to use the machine learning approach in its simplest and transparent formulation with single hidden layer \cite{Melko1} to explore complex non-collinear magnetic phases of technological importance? In this respect topologically-protected magnetic skyrmions \cite{Bogdanov1, Bogdanov2, skyrmion1, skyrmion2, skyrmion3} and spin spiral states are the first candidates for such a consideration, since they can be used for creating novel magnetic memory devices \cite{expdisk}.  Numerous experimental studies revealed skyrmion state in metallic ferromagnets with Dzyaloshinskii-Moriya interaction such as FeGe \cite{FeGe1,FeGe2}, Fe monolayer on Ir(111) \cite{FeIr}, MnGe \cite{MnGe}, Fe$_x$Co$_{1-x}$Si \cite{Yu} in a narrow range of the external parameters, magnetic fields and temperatures. The experimental phase diagrams of these materials \cite{Yu} contain significant transitional areas between different phases, which raises the problem of the precise definition of the skyrmion and spin spiral phase boundaries. 

Here we show that a standard feed-forward network (FFN) can be used efficiently for supervised learning on topologically-protected magnetic skyrmion states and spin spirals originated from the spin-orbit coupling. Fig.\ref{neural network} illustrates the idea of our approach. A non-collinear magnetic configuration obtained from the Monte Carlo simulations describing a two-dimensional ferromagnet with Dzyaloshinskii-Moriya interaction (Fig.\ref{neural network} a) is projected on the $z$ axis (Fig.\ref{neural network} b). This $z$-projected magnetic structure is considered as input for the single-hidden-layer network (Fig.\ref{neural network} c). Having trained such a network on a limited set of the configurations belonging to pure ferromagnetic, skyrmion and spiral states on the square lattice we were able to recognize the states from completely different parts of the phase diagram, including transitional areas between different phases. Moreover, we found that the trained network can classify the data collected for triangular lattice, which demonstrates universality of our approach. Another important result of our work is the demonstration of a high classification performance achieved with a nearest centroid method. Being one of the simplest machine learning techniques the centroid classifier nevertheless shows very accurate results in case of unseen data on skyrmion and spin spiral configurations.    

\begin{figure}[b] 
\center 
\includegraphics[width=\columnwidth]{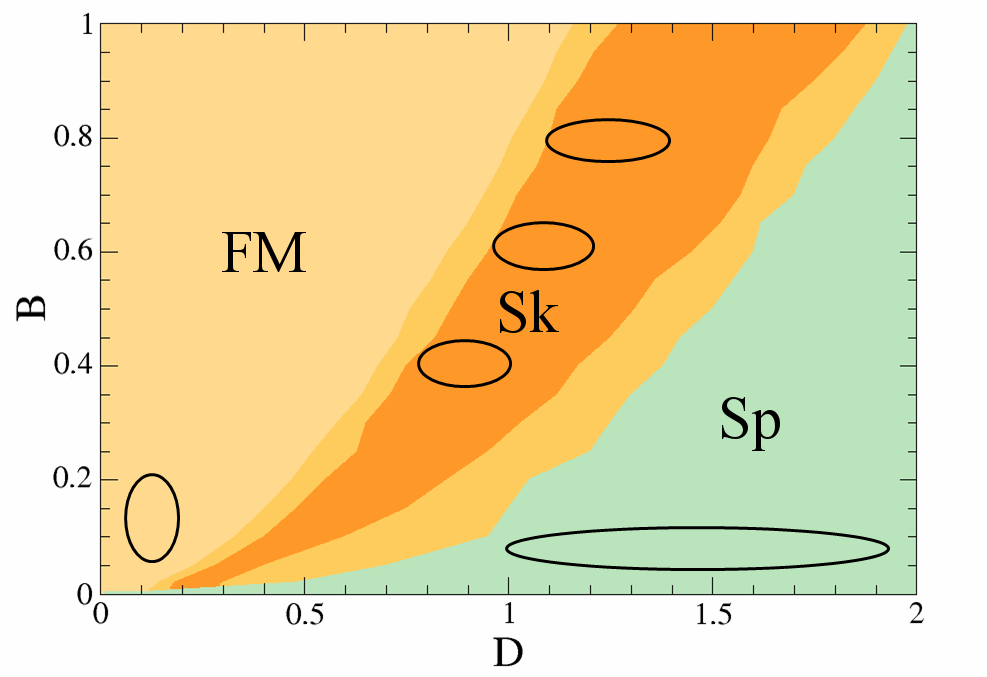} 
\caption{Phase diagram in terms of Dzyaloshinskii-Moriya interaction and magnetic field. The abbreviation Sk, FM and Sp denote skyrmion lattice state, ferromagnetic and spin spiral state, respectively. The phase diagram was obtained at $T=0.02$. All the parameters are given in units of $J$. Black ovals denote the phase areas used for supervised learning.}
\label{family} 
\end{figure}

\section{Model and Method} 
In our study to simulate the topological magnetic excitations we used the following spin Hamiltonian on the 48$\times$48 square lattice:
\begin{equation}\label{Ham}
\begin{split}
H=- \sum_{i<j}J_{ij}{\bf S}_i{\bf S}_j-\sum_{i<j}{\bf D}_{ij}[{\bf S}_i\times{\bf S}_{j}]-  \sum_i B S_i^z,
\end{split}
\end{equation}
where $J_{ij}$ and ${\bf D}_{ij}$ are the isotropic exchange interaction and Dzyaloshinskii-Moriya vector, respectively. ${\bf S}_{i}$ is a unit vector along the direction of the $i$th spin and $B$ denotes the $z$-oriented magnetic field. We take into account the only interactions between nearest neighbours. The isotropic exchange interaction is positive in our simulations, which corresponds to the ferromagnetic case. The symmetry of the Dzyaloshinskii-Moriya vectors is of $C_{4v}$ type, DMI has an in-plane orientation and perpendicular to the corresponding inter-site radius vector. The Hamiltonian was solved by using the classical Monte Carlo approach. The spin update scheme is based on the Metropolis algorithm. The systems in question are gradually (200 temperature steps) cooled down from high temperatures (${\rm T}\sim 3J$) to the required temperature. Each temperature step run consists of $1.5\times10^{6}$ Monte Carlo steps.

To identify the different magnetic phases of the spin Hamiltonian, Eq.(\ref{Ham}) we calculated spin-spin correlations functions \cite{Danis}, topological charges \cite{Berg} (the corresponding expressions are presented in Appendix A) and visualized a number of the spin configurations from each simulation.
By using such information a neural network was trained as described below.

\section{Neural Network}
In our study we employ a standard network architecture that is one-layer feed forward network presented in Fig.\ref{neural network} c. It consists of one hidden layer of sigmoid activation neurons and three output sigmoid neurons that activate depending on the particular magnetic phase. For the training set we generated 1000 configurations for each of ferromagnetic, skyrmion and spiral states corresponding to the areas marked in Fig.\ref{family}. In these simulations we fixed $J=1$ and used a uniform distribution for magnetic field and Dzyaloshinskii-Moriya interaction. The simulation temperatures were taken in the range $T\in[0.02,0.1]$ in units of isotropic exchange interaction. Moreover, we generated 1000 configurations belonging to paramagnetic phase at high temperatures ($T \sim 10J$) and added them to the training set. For these paramagnetic configurations the ground-truth labels of all the output neurons were set to zero.  

The main challenge in a machine learning for classification of magnetic phases is how to relate the states of the input neurons of the network to the particular magnetic configuration. As it was shown in Ref.\onlinecite{Melko1} in the case of the Ising model with $S^{z} = \pm 1$ there is one-to-one correspondence between the neuron values in the input layer and spins of the particular configurations. On the other hand, for the XY model solutions characterized by in-plane non-collinear magnetic states the authors of Ref.\onlinecite{Melko2} used the angle values determining the in-plane orientations of the spins. 

In the case of the non-collinear magnetic configurations the situation is more complicated, since the orientation of a spin can not be described by single angle value. However, one can make use of that skyrmions are characterized by a typical profile, the core and background spins of a skyrmion align anti-parallel and parallel to the applied magnetic field (Fig.\ref{neural network} b), respectively. It means that the skyrmion excitation can be detected by analyzing the $z$ components of the spins \cite{Stamps}. We use this fact to realize our neural network approach, the values of the input neurons are equal to the $z$ components of the spins obtained from Monte Carlo simulations of Eq.(\ref{Ham}). As we will show below, such an approach also works well in the case of the spin spiral and ferromagnetic phases.

The network was trained to minimize the error function that is a standard Mean Squared Error (MSE) function. Weights of neurons were adjusted by means of back-propagation method. Details of the learning process are given in Appendix B. The network was trained with different numbers of hidden neurons from 8 to 128. According to our simulations the network with 64 hidden neurons gives reliable results on the phases recognition. We found that further increase of hidden neurons number for considered case leads to decrease in recognition quality. Thus, the total number of adjustable parameters are $64L^2+192$, which is much smaller than in the previous work \cite{Melko2}.

{\it Phase diagram.} The developed neural network approach was used for construction of the phase diagram of the spin model, Eq.(\ref{Ham}) on the square lattice. To do this we used a grid of 625 points on the temperature-magnetic field plane. For each point the values of the neural network output neurons were averaged over 10 Monte Carlo runs. Thus the total number of the Monte Carlo calculations was equal to 6250. 

\begin{figure}[t] 
\center 
\includegraphics[width=\columnwidth]{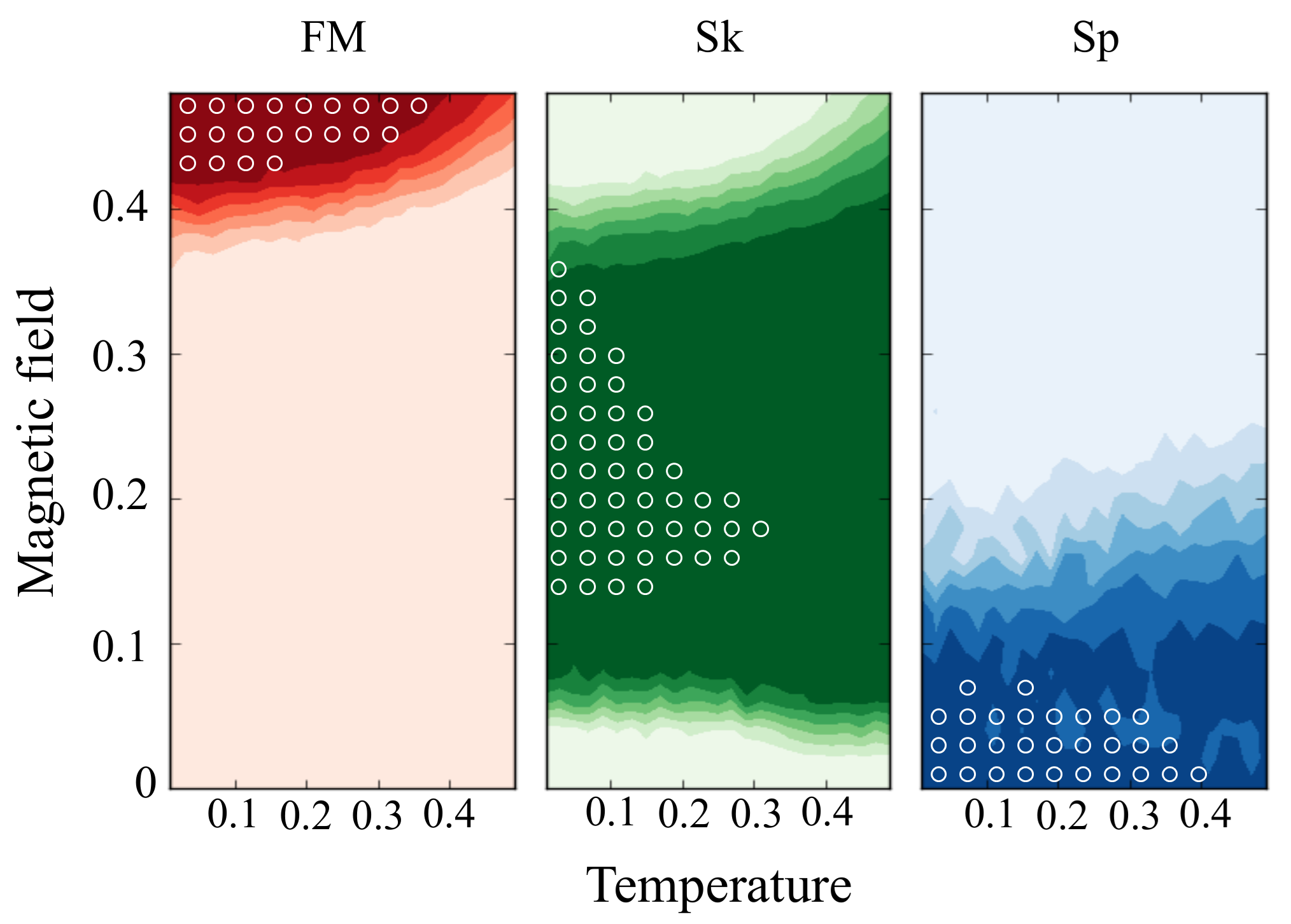} 
\caption{Phase triptych obtained by using the neural network with 64 hidden neurons. Color intensities indicate the values of the output neurons for different phases, dark and light colors correspond to 1 and 0, respectively. The Dzyaloshinskii-Moriya interaction was chosen to be $D=0.72$. All the parameters are given in units of $J$. White circles denote the phases boundaries defined with the spin structure factors.}
\label{phase} 
\end{figure}

From Fig.\ref{phase} one can see that the trained network can successfully recognize all the phases of interest at low temperature, which follows from a comparison with the boundaries obtained by calculating the structure factor (white circles). It is worth mentioning that we obtained a large value of skyrmion number ($Q>15$) for the parameters corresponding to the dark green area in Fig.\ref{phase}. Importantly, it is possible to perform a composition analysis of the transitional areas between different phases. For each point of the phase diagram one can define the values of the output neurons that indicate the contributions of the phases. It gives us opportunity to solve the complex problem of the definition of the phase boundaries and quantitatively characterize the transitional areas between different phases \cite{bimeron1,bimeron2}. 

\begin{figure}[b]
	\includegraphics[width=\columnwidth]{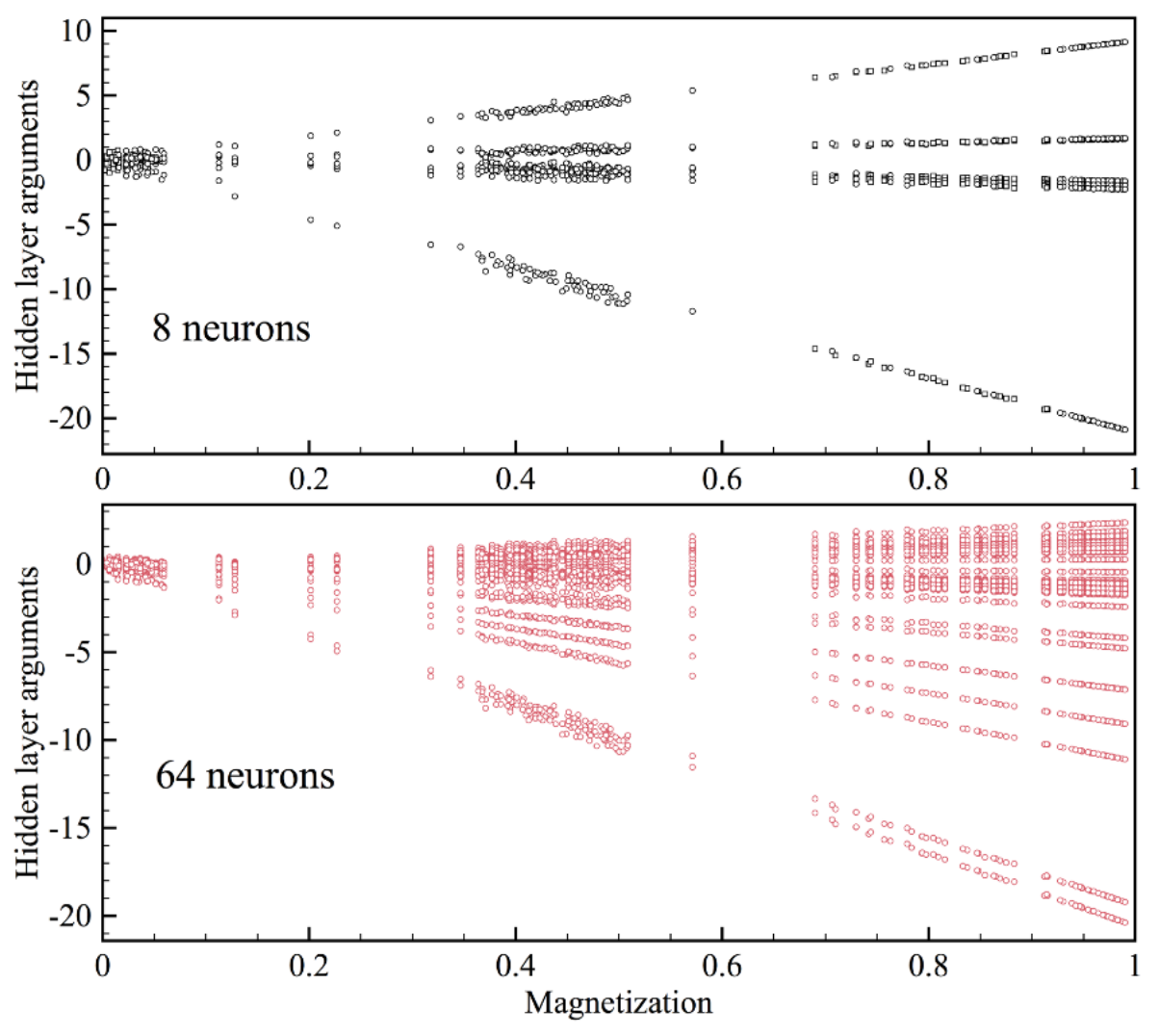} 
	\caption{Hidden layer arguments as a function of the $z$-oriented magnetization of the simulated spin configurations. (Top) 8-hidden-neuron network. (Bottom) 64-hidden-neuron network.}
	\label{fig:Melko} 
\end{figure}

\begin{figure}[b]
	\includegraphics[width=\columnwidth]{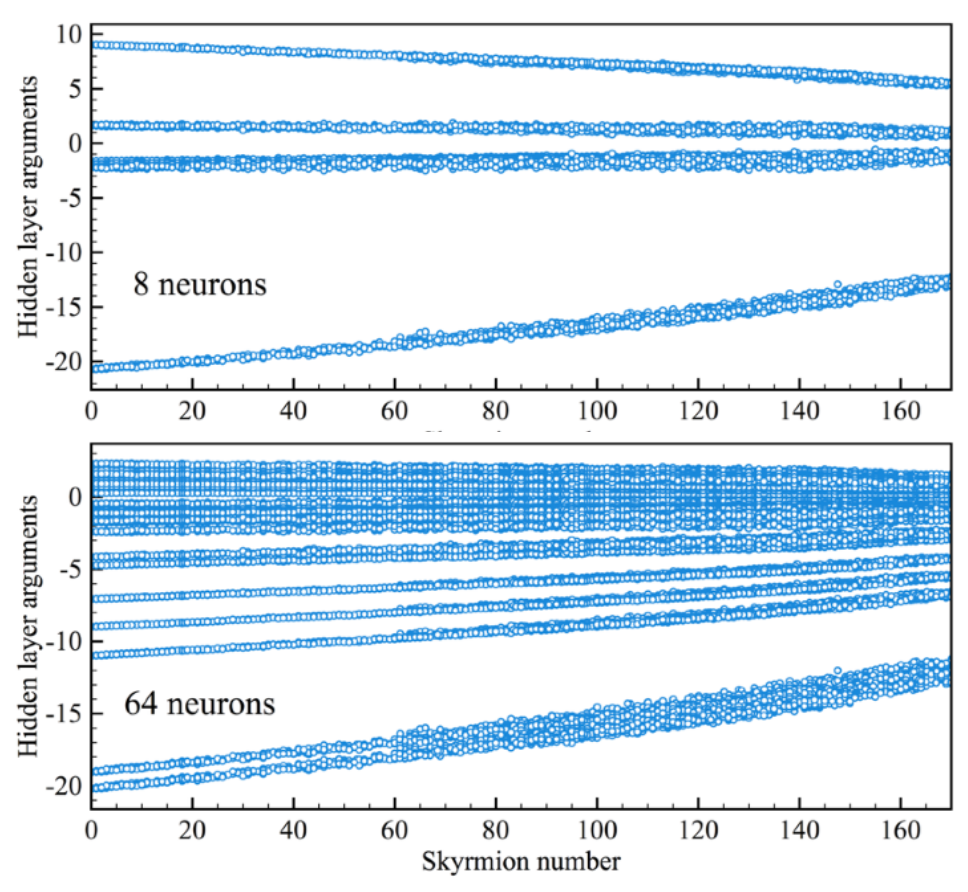} 
	\caption{Hidden layer arguments as a function of the skyrmion number of the simulated spin configurations.}
	\label{fig:Melko2} 
\end{figure}

\begin{figure}[t] 
\center 
\includegraphics[width=\columnwidth]{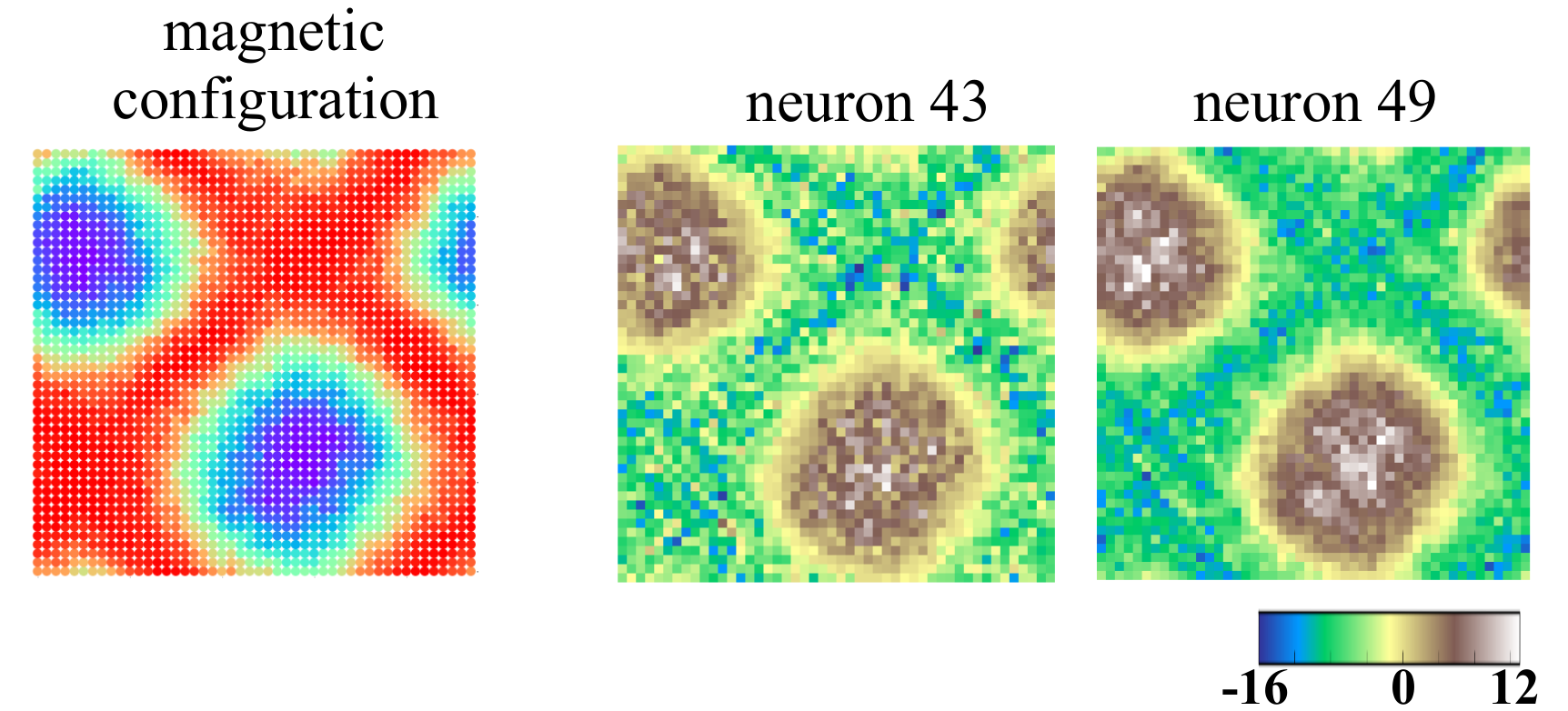} 
\caption{(Left panel) $z$-projection of the skyrmion magnetic configuration obtained with the parameters $J=1, D=0.2, T=0.02$ and $B=0.02$. (Right panel) Visualization of the arguments of the specific hidden layer neurons.}
\label{big_skyrmion} 
\end{figure}

{\it Analysis of the classification process.}
The results of the previous neural-network-based studies \cite{Melko1,Melko2,Troyer,Wessel} stand new fundamental questions on how a network learns different phases of matter. It was shown in Ref.\onlinecite{Melko1} that identification of the Ising model states is related to difference in total magnetization of the spin configurations belonging to different phases. In our case such an explanation can be also used, since the phases we simulated are characterized by different magnetizations. The magnetization per spin, $m(x) = \frac{1}{N} \sum_{i}^N S^z_{i}$ in the training set is in the range [0.91, 0.99], [0.38, 0.53] and [0, 0.03] for ferromagnetic, skyrmion and spin spiral phases, respectively. At the same time, the test sets include pure spin configurations that are characterized by wider ranges of the average magnetization: [0.84,  0.99] (ferromagnetic), [0.33, 0.69] (skyrmion) and [0, 0.07] (spin spiral). In agreement with Ref.\onlinecite{Melko1} we obtain that the components of the vector $W x$ (here $W$ is the weights matrix between input and hidden layers) become linear functions of the magnetization $m(x)$ (Fig. \ref{fig:Melko}). However, in our case the increase of the number of the hidden neurons leads to a larger number of the neuron categories, which may mean that the magnetization is not the only parameter the network uses for recognition.

\begin{figure}[b] 
\center 
\includegraphics[width=\columnwidth]{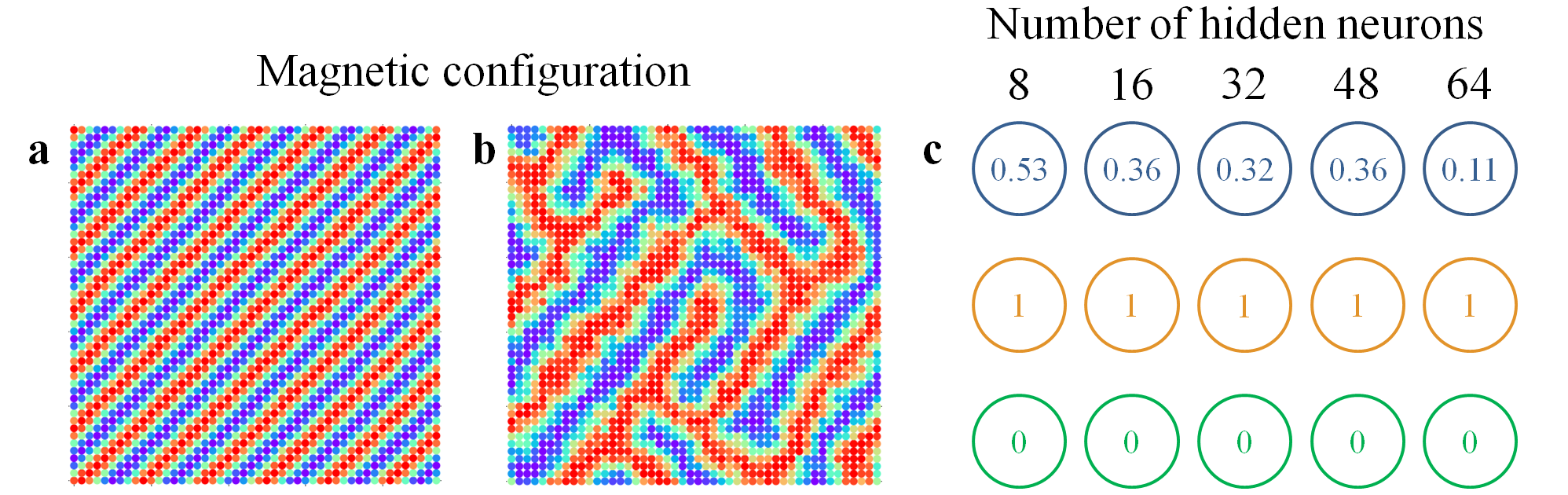} 
\caption{(a) Example of spiral state ($D$=1.4, $B$=0.02, $T$=0.05, $J$=1) used for training the network. (b) Example of a complex spiral configuration from a test set obtained with $D$=0.72, $B$=0.03, $T$=0.22, $J$=1. (c) The output neurons values in the case of configuration (b) depending on the number of the hidden neurons.  Numbers in blue, orange and green circles correspond to values of skyrmion, spiral and FM outputs, respectively.}
\label{labyrinth} 
\end{figure}

Since our focus in this study is on the skyrmion phase recognition, we have also investigated the dependence of the hidden neurons arguments on the topological charge. For that the pure DMI model with zero isotropic exchange interaction was simulated with varying magnetic field. It gives us opportunity to produce 2000 magnetic configurations characterized by completely different skyrmion numbers (from 0 to 170) with the same system size. These results are presented in Fig.\ref{fig:Melko2}. All the neurons can be divided into two categories. The first one corresponds to the neurons with argument values that are close to zero and not sensitive to the topological charge. Another one depends on the skyrmion number of the particular magnetic configuration. 

To understand the neural network functioning one can also visualize hidden layer neurons. By the example of the configuration with big skyrmions presented in Fig.\ref{big_skyrmion} we performed such an analysis. Importantly, the size of the skyrmions in the training data set does not exceed $10a$, where $a$ is the lattice constant, but we found, that the trained neural network correctly classifies the configurations with skyrmions of much larger diameter. Indeed, the diameter of the skyrmion in Fig.\ref{big_skyrmion} is about $35a$ and such a skyrmion state is uniquely recognized by the neural network even with 8 hidden neurons. 

Figure~\ref{big_skyrmion} gives two-dimensional representation of two hidden neurons arguments that are the weights matrix multiplied by spin $z$ components corresponding to the magnetic configuration. The maximal and minimal intensities of the core and background areas of the skyrmions are different for these neurons. Nevertheless, one can easily recognize the original skyrmion structure. The visualization of the neural network weights by themselves does not give any useful information about network functioning. 

As a hard test for our neural-network approach we generated 300 high temperature spiral configurations ($T\in[0.18;0.26], D=0.72, B=0.03$). A typical example of such configurations is presented in Fig.\ref{labyrinth} (b). It is of labyrinth type and consists of the broken spin spirals that are distorted due to the temperature effects. Importantly, the training set contains only ideal spin spirals presented in Fig.\ref{labyrinth} (a). One can see that increase of the number of the hidden neurons leads to a decrease in the value of the output neuron corresponding to the skyrmion phase that provides a more accurate phase separation. Having analyzed this test set, we found that total number of clearly recognized configurations increased from $40\%$ to $75\%$ with using 8 and 64 hidden neurons, respectively.

\begin{figure}[t]
	\includegraphics[width=\columnwidth]{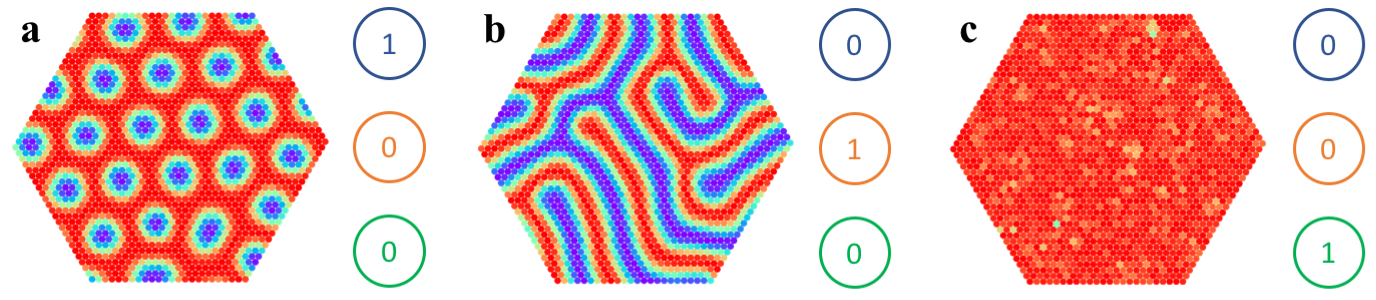} 
	\caption{Examples of skyrmion (a), spin spiral (b) and ferromagnetic (c) configurations stabilized on the triangular lattice and recognized with the neural network trained on square lattice data. Numbers in blue, orange and green circles correspond to values of skyrmion, spiral and ferromagnetic outputs, respectively.}
	\label{fig:deep_dream_triangular} 
\end{figure}

{\it Variation of the lattice structure.}---The next step of our investigation is to examine the network trained on the square lattice magnetic configurations for recognizing the phases of the spin Hamiltonian on the triangular lattice. For that we solved Eq.(\ref{Ham}) with DMI of $C_{3v}$ symmetry and generated magnetic configurations belonging to skyrmion, spin spiral and ferromagnetic phases as well as their mixtures. Fig.\ref{fig:deep_dream_triangular} gives the corresponding examples. For preparation of the test configurations we have solved spin Hamiltonian, Eq.(1) on the triangular lattice 48$\times$48 with periodic boundary conditions. The supercell of the rhombic shape was replicated. A square area of 48$\times$48 spins cropped from the replicated lattice was used to define the values of the neural network input neurons.

It was found that the trained network classifies the skyrmion and ferromagnetic triangular-lattice configurations with high precision. In the case of the spin spiral states the classification accuracy is low, since such magnetic configurations (typical example is presented in (Fig.\ref{fig:deep_dream_triangular} b) strongly differ from those we used in the training set (Fig.\ref{labyrinth} a).

\section{Minimum distance (nearest centroid) classification}
As it was shown in the previous section the neural network approach paves the way to explore the magnetic phase diagram of non-collinear magnets including mixed states such as spin-spiral-skyrmion and skyrmion-ferromagnetic. At the same time the problem, when a particular state should be assigned to only one magnetic phase, can be solved with a much simpler method.   

\begin{figure}[t]
	\includegraphics[width=\columnwidth]{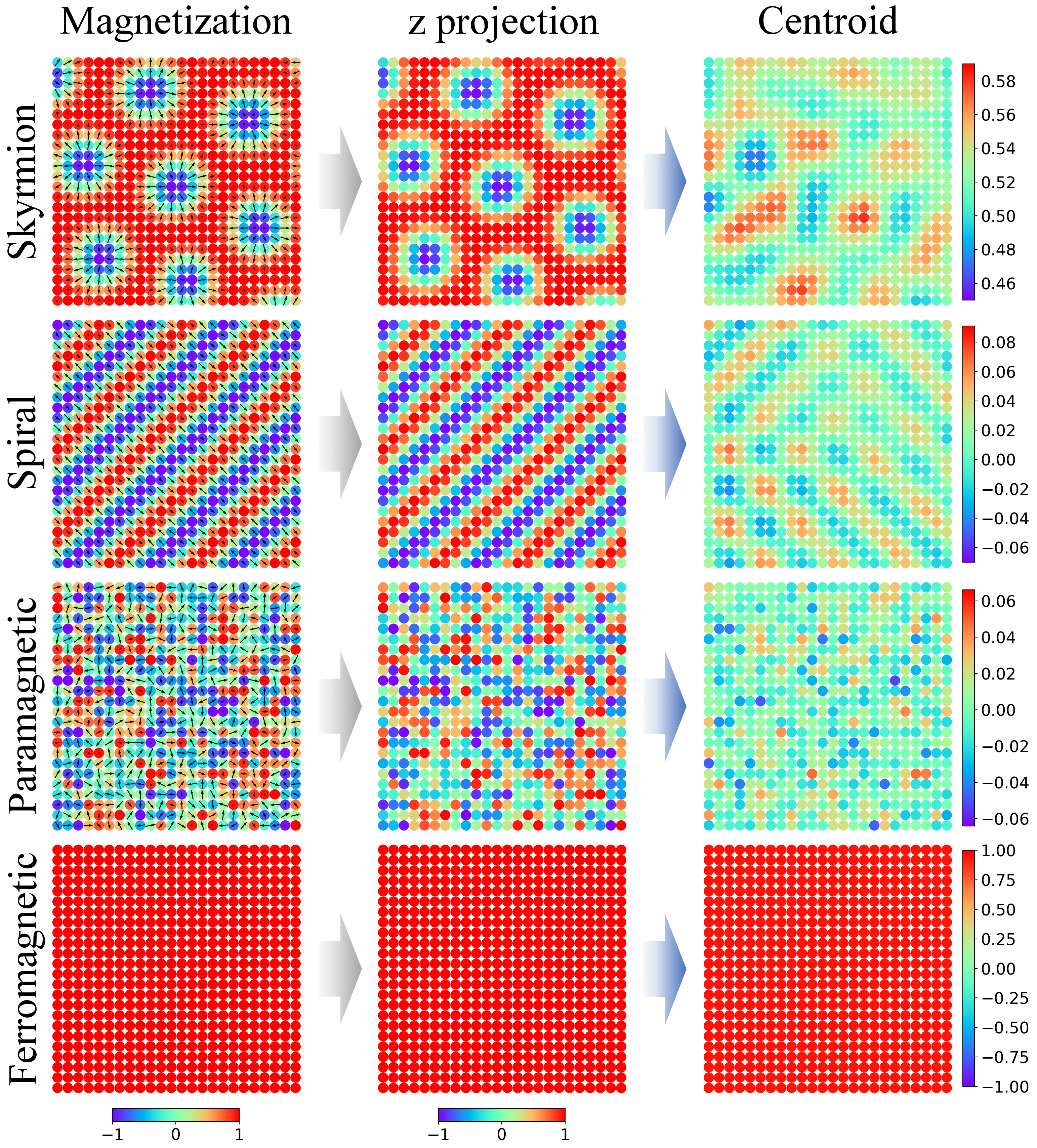} 
	\caption{Comparison of centroids for different phases. Left panels are the examples of skyrmion, spin-spiral, paramagnetic and ferromagnetic configurations. The arrows denote in-plane orientations of the magnetic moments. Middle panels are the corresponding z projections of the example magnetic configurations. Right panels represent two-dimensional visualisations of the centroids calculated with Eq.\ref{centroids} for training data sets.}
	\label{fig:nearest_centroids} 
\end{figure}

In this section we utilize the nearest centroid classification method as implemented in scikit-learn python package~\cite{kNN}. Figure~\ref{fig:nearest_centroids} shows the overall process of the classifier training. As in the case of FFN we perform data preprocessing by projecting local magnetization vectors on $z$ axis. The next step is to  calculate mean data values for each class $\alpha$ (magnetic phase) in the training data set. These mean values are called centroids and given by

\begin{equation}
 \braket{\mathbf{X}}_\alpha = \frac{1}{M}\sum_{k=1}^M\mathbf{X}^{(k)}_\alpha, \label{centroids}
\end{equation}
where $ \mathbf{X}^{(k)}_{\alpha} = \{S_{z1}^{(k)}, S_{z2}^{(k)}, \ldots, S_{zN}^{(k)}\} $ is a vector formed from $z$ components of local magnetization $ S_{zi}^{(k)} $ for $ k $th magnetic configuration and $\alpha = $ FM, PM, Sk, Sp denotes a phase. Thus, one can identify the phase $\alpha_{test}$ of a magnetic configuration $\mathbf{X}^{test}$ by determining the minimum distance from it to the centroid of each class (magnetic phase):

\begin{equation}
    d = {\rm min}_{\alpha}\left\{||{\braket{\mathbf{X}}_{\alpha{}} - \mathbf{X}^{test}}||\right\},\label{norm}
\end{equation}
where $||...||$ means norm of a high-dimensional vector.

As in the case of the neural network for training of nearest centroids classifier we used the same set comprising 4000 square lattice magnetic configurations. Figure~\ref{fig:nearest_centroids} (right panels) gives two-dimensional visualizations for the calculated centroids of different magnetic phases. As one can expect, the maximal and minimal centroid intensities are connected to average  magnetization per spin for each phase. However, each centroid has distinct magnetic pattern inherent to the corresponding phase. For example, the average magnetization values (per spin) of spin spiral and paramagnetic phases are close to zero, but centroid of spiral phase preserves the ordering, whereas the mean of paramagnetic phase configurations stays disordered. This feature allows the method to distinguish PM and spin spiral phases.

\begin{figure}[t]
	\includegraphics[width=\columnwidth]{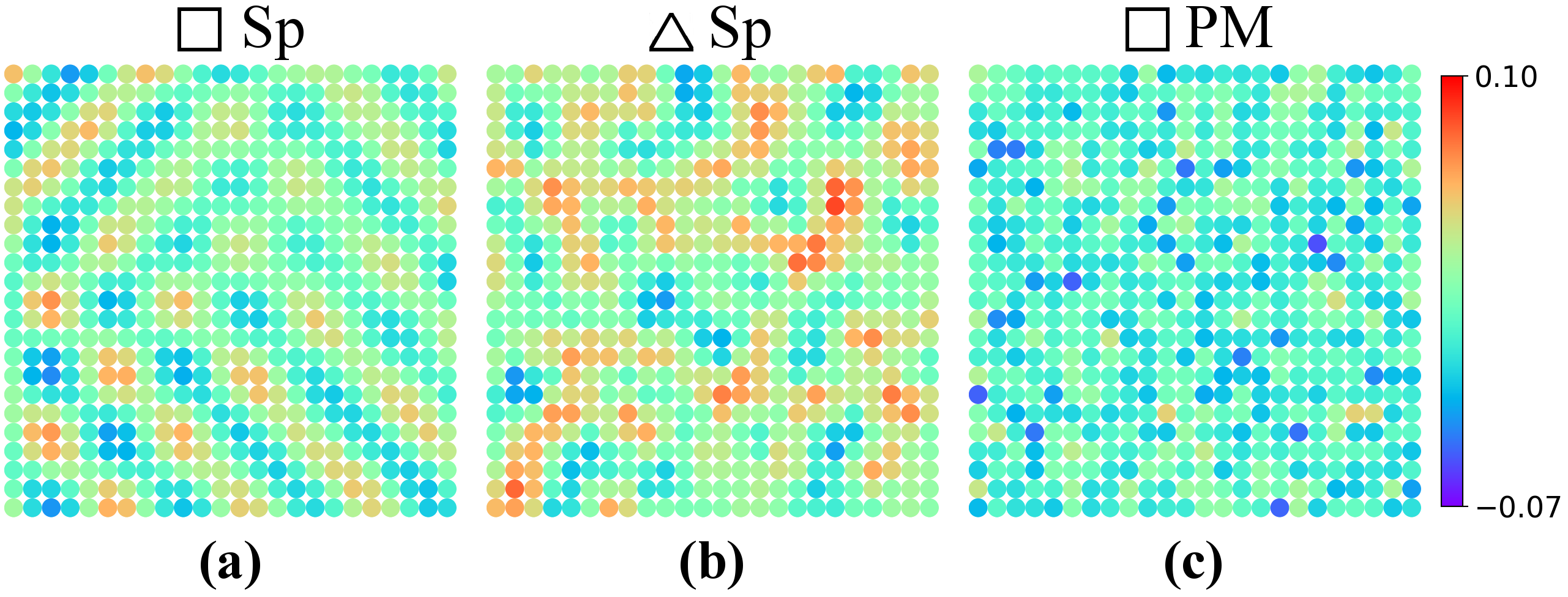} 
	\caption{Comparison of centroids calculated with (b) test set configurations on the triangular lattice and (a, c) training set on the square lattice.}
	\label{fig:square_and_triangular_centroids} 
\end{figure}

The next important step is to estimate the performance of the algorithm on the unseen data, such as big skyrmions (Fig.\ref{big_skyrmion}), high-temperature spin spirals (Fig.\ref{labyrinth}) and triangular lattice configurations (Fig.\ref{fig:deep_dream_triangular}). The results of classification are presented in Table~\ref{tab:ml_comparison}. It was found, that both neural network and nearest centroids classifier show comparable recognition accuracy for big skyrmions (94\% and 100\%), high temperature spirals (75\% and 78\%), ferromagnetic (both 100\%) and skyrmion configurations on triangular lattice (91\% and 100\%). Finally, the nearest centroid algorithm demonstrates excellent performance for classification of paramagnetic phase (90\%). The centroid classifier shows slightly better classification of 880 spin spiral configurations stabilized on triangular lattice (54\%) than the neural network one, but recognition accuracy is still low. This can be due to different topology of underlying spirals structure for triangular and square lattices. Indeed, Fig.~\ref{fig:square_and_triangular_centroids} shows that the centroid patterns for spin spirals stabilized on triangular and square lattices are completely different. One can also note that the calculated distances $ || \braket{X}_\alpha - \braket{X}_{test} || $ from triangular lattice spiral set centroid to square lattice spiral and paramagnetic sets centroids are comparable (1.7 and 1.8, respectively), thus explaining the part of triangular spiral configurations recognized as paramagnetic.

\begin{table}[t]
    \centering
    \caption{Comparison of different ML classifiers trained with the same data set. Testing sets include 100 big skyrmions (Big Sk), 300 high temperature spin spirals (HT Sp) and data set for triangular lattice ($\Delta$).}\label{tab:ml_comparison}
    \begin{tabular}{rccccc}
    \hline
    \hline
        \multicolumn{2}{c}{data set} &\, & FFN,\,\%\, & mean,\,\%\, & $k$-NN,\,\%\, \\\hline
        Big & Sk &\, & 94 & 100 & 0  \\
        HT & Sp &\, & 75 & 78 & 9 \\
        $\Delta$ & FM &\, & 100 & 100 & 100 \\
        $\Delta$ & Sp &\, & 40 & 54 & 25 \\
        $\Delta$ & Sk &\, & 91 & 100 & 48 \\
        $\Delta$ & PM &\, & 37 & 90 & 100\\\hline \hline
    \end{tabular}
\end{table}


\section{$k$-nearest neighbor classifier} 
The choice of the best classifier algorithm is non-trivial task and highly depends on the nature of classified data and the purpose of classification. Often one needs to test a number of approaches to find appropriate one. Here, we present results obtained for $k$-nearest neighbor method, which is widely used for classification tasks.

\begin{figure}[b]
	\includegraphics[width=\columnwidth]{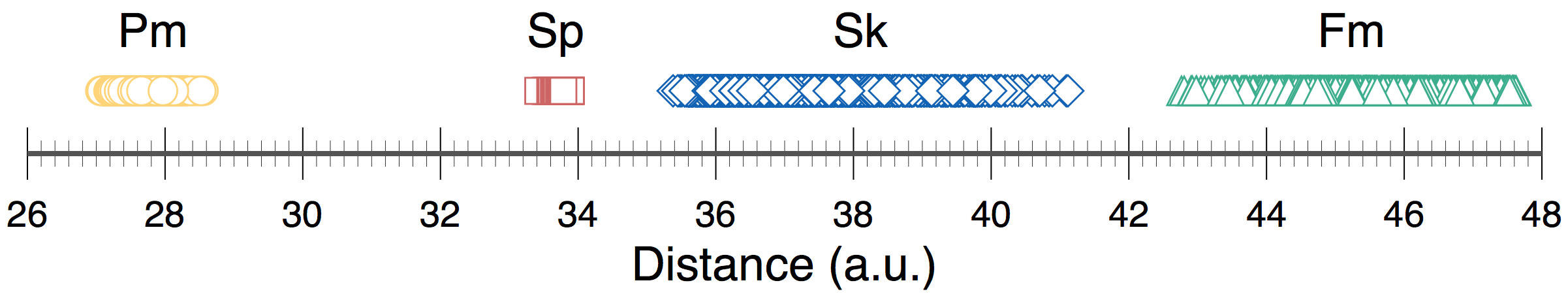} 
	\caption{One-dimensional visualization of training set comprising 4000 magnetic configurations. Yellow circles, red squares, blue diamonds and green triangles denote magnetic configurations belonging to paramagnetic, spiral, skyrmion and ferromagnetic phases, respectively. They are distributed with respect to the distance from origin in 2304-dimensional space (48$\times$48 spins in total for each configuration).}
	\label{knn} 
\end{figure}

Like for the nearest centroids classifier, in $k$-NN method a magnetic configuration will be assigned to the specific class of the magnetic configurations from the test set by using the distance metric, however, the classification is now based on the closest $k$ neighbors in the feature space that is the space of the magnetization vector elements. It is known that best choice of parameter $k$ is highly dependent on the nature of data. We found that the classification scheme based on three nearest neighbors ($k$ = 3) shows best results for our calculations. Figure~\ref{knn} gives one-dimensional representation of the training set within the $k$-NN method. There is a clear separation of the magnetic configurations belonging to the different phases. Since the training in $k$-NN algorithm is simply storing of the magnetization vectors, the neural network approach works slower at this stage. 

To estimate the performance of the $k$-NN algorithm we carried out classification for our test data sets presented in Table~\ref{tab:ml_comparison} with this method. It was found $k$-NN method improperly assigns the big skyrmions to ferromagnetic phase configurations whereas neural network correctly classifies 94\,\% of such skyrmions. Then only 9\,\% of 300 high-temperature spin spiral configurations of the labyrinth type were correctly classified. At the same time, the neural network approach demonstrates 75\,\% accuracy for this test. Both $k$-NN and neural network methods show 100\,\% classification results in the case of the ferromagnetic configurations (880 in total) stabilized on the triangular lattice. In turn, neural network clearly surpasses $k$-NN for skyrmion states (880 in total) on the triangular lattice, 91\,\% against 48\,\%. Nevertheless, $k$-NN classifier correctly recognizes all paramagnetic configurations.

\section{Conclusions} 
We have developed a neural-network-based approach for recognition magnetic phases of two-dimensional ferromagnets with Dzyaloshinskii-Moriya interaction in wide ranges of magnetic fields and temperatures. One needs to generate a limited set of magnetic configurations ($\sim$ 4000 in total) to train the network. It facilitates the construction of the phase diagram of the system in question during the Monte Carlo sampling. Complex and mixed ferromagnetic-skyrmion and skyrmion-spin-spiral configurations can be quantitatively described, which was not possible before. The calculations for spin Hamiltonians on the 128$\times$128 square lattice also demonstrated high accuracy in classification of the magnetic phases. We have shown that the method does not sensitive to the particular lattice structure used for training. By construction the network approach allows one to recognize the skyrmions of different types (Bloch and N\'eel). It can be used for on-the-fly classification of the skyrmion magnetic configurations observed in experiments. We have also utilized other widely used methods of machine learning classification and shown that the proposed method demonstrates comparable performance with nearest centroid classification method (except for paramagnetic phase) and totally surpasses the k-nearest neighbors method.

\section{Acknowledgements} 
We thank Andrey Bagrov, Fr\'{e}d\'{e}ric Mila and Jeanne Colbois for fruitful discussions.
This work was supported by the Russian Science Foundation Grant 18-12-00185.

\appendix
\section{Problem demonstration}
The aim of this section is to demonstrate the complexity of the magnetic phase classification problem by the example of the skyrmionic materials. In our previous work~\cite{bimeron2} we have shown that there are five stable phases in a system described by the spin Hamiltonian, Eq.\eqref{Ham} which can be uniquely identified at the low temperature by calculation of the spin structure factors and the skyrmion number (topological charge).
The expressions for spin structure factors are given by
\begin{equation}\label{hi1}
\chi_\parallel(\textbf{q})=\frac{1}{N}\left\langle\left|\sum_{i}S^z_ie^{-i\textbf{q}\textbf{r}_i}\right|^2\right\rangle,
\end{equation}

\begin{equation}\label{hi2}
\chi_\perp(\textbf{q})=\frac{1}{N}\left\langle\left|\sum_{i}S^x_ie^{-i\textbf{q}\textbf{r}_i}\right|^2+\left|\sum_{i}S^y_ie^{-i\textbf{q}\textbf{r}_i}\right|^2\right\rangle,
\end{equation}
where $\textbf{q}$ is the reciprocal space vector, $S_i^\alpha$ ($\alpha = (x,y,z)$) is the projection of the $i$th spin and $\textbf{r}_i$ is the radius vector for the $i$th site.

\begin{figure}[b] 
	\center 
	\includegraphics[width=\columnwidth]{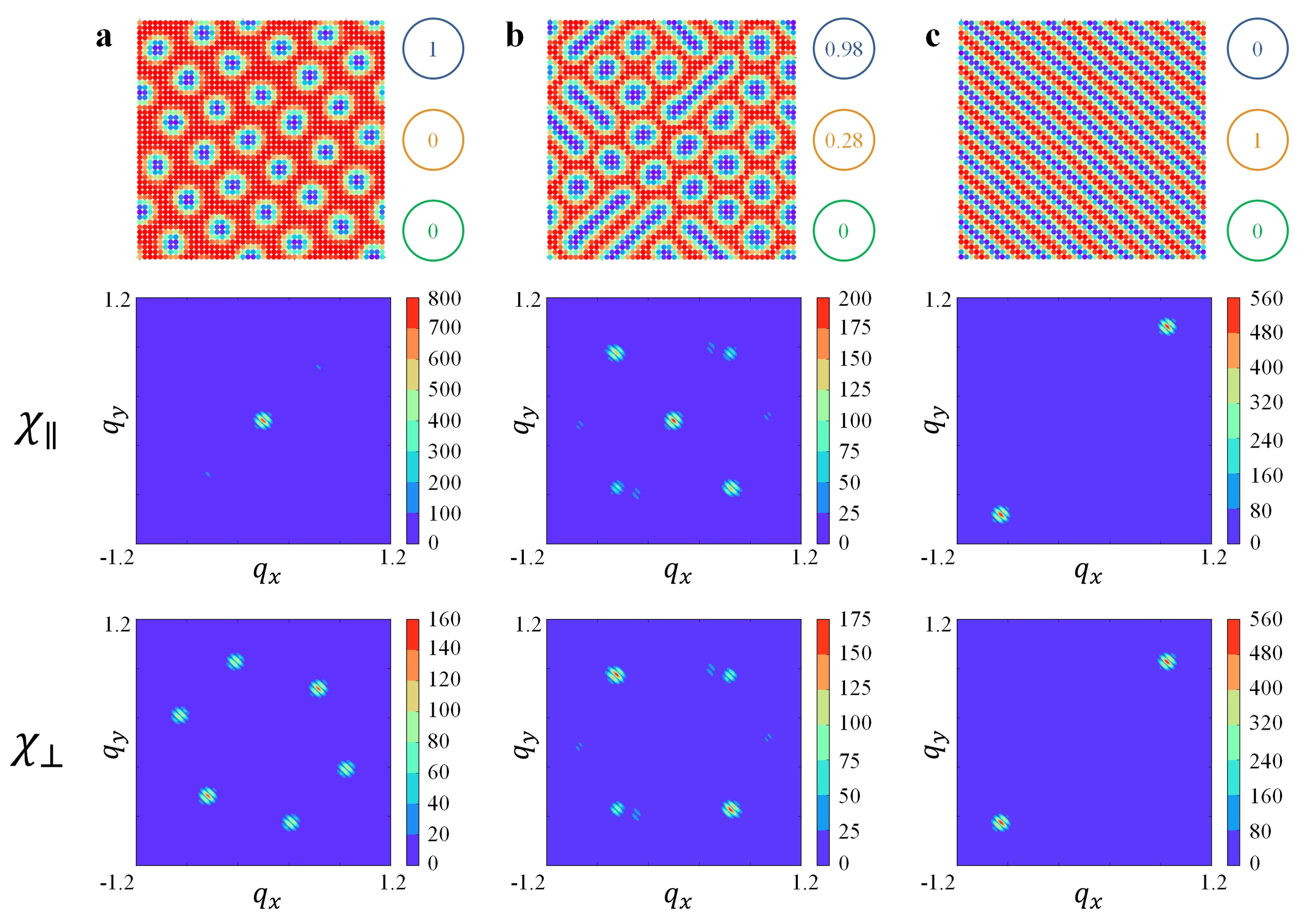} 
	\caption{Examples of pure skyrmion (a), mixed skyrmion-bimeron (b) and pure spiral (c) magnetic configurations obtained at the low temperature ($T=0.02J$) and corresponding spin-structure factors. Numbers in blue, orange and green circles correspond to values of skyrmion, spiral and FM outputs, respectively. Skyrmion numbers of these configurations are equal to 32, 28 and 0 from left to right.}
	\label{str1} 
\end{figure}
In turn, the topological charge is defined in the following way
\begin{equation}\label{Q}
Q=\frac{1}{4\pi}\sum_{l}A_{l},
\end{equation}
where $A_l$ is the solid angle subtended by three spins located at the vertices of an elementary triangle $l$,
\begin{equation}\label{A_l}
A_l=2\arccos\left(\frac{1+\textbf{S}_i\!\cdot\textbf{S}_j+\textbf{S}_j\!\cdot\textbf{S}_k+\textbf{S}_k\!\cdot\textbf{S}_i}{\sqrt[]{2(1+\textbf{S}_i\!\cdot\textbf{S}_j)(1+\textbf{S}_j\!\cdot\textbf{S}_k)(1+\textbf{S}_k\!\cdot\textbf{S}_i)}}\right).
\end{equation}
The sign of $A_l$ in Eq.~\eqref{Q} is determined as $sign(A_l)=sign(\textbf{S}_i\cdot[\textbf{S}_j\times\textbf{S}_k])$. Importantly, we do not consider the exceptional configurations for which
\begin{align}
&{\bf S}_{i} \cdot [{\bf S}_{j} \times {\bf S}_{k}] = 0, \\
&1+ {\bf S}_{i} \!\cdot {\bf S}_{j} + {\bf S}_{j} \!\cdot {\bf S}_{k} + {\bf S}_{k} \!\cdot {\bf S}_{i} \le 0. \notag
\end{align} 

\begin{figure}[t] 
	\center 
	\includegraphics[width=\columnwidth]{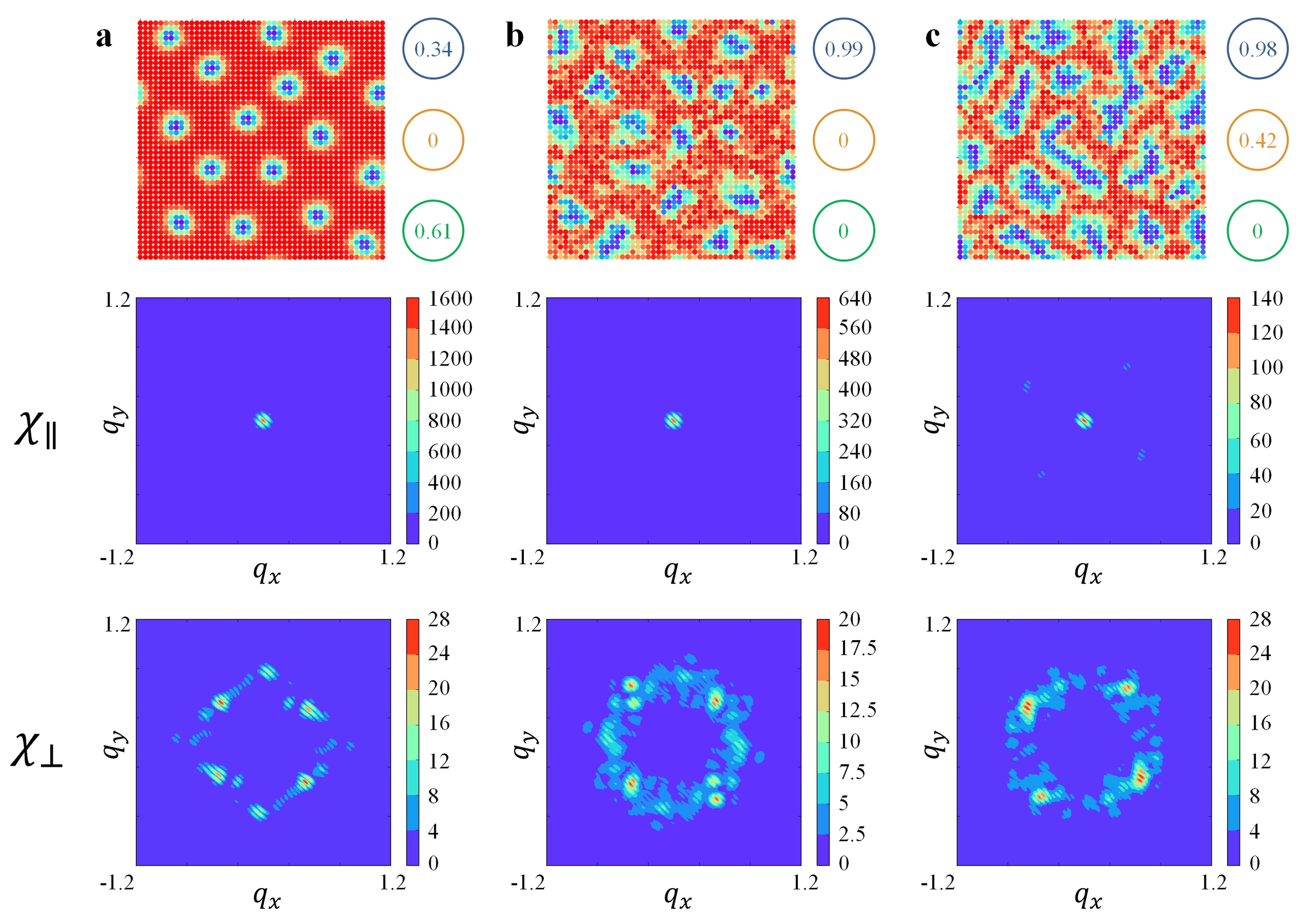} 
	\caption{Examples of non-periodic skyrmions (a) obtained at the low temperature ($T=0.02J$), pure skirmions (b) and mixed skyrmion-bimeron (c) magnetic configurations obtained at the high temperature ($T=0.4J$) and corresponding spin-structure factors. Numbers in blue, orange and green circles correspond to values of skyrmion, spiral and FM outputs, respectively. Skyrmion numbers of the presented configurations are equal to 15, 19 and 15 from left to right.}
	\label{str2} 
\end{figure}
  
Figure~\ref{str1} gives the examples of the most interesting phases. As can be seen all of them are recognized correctly by the network. However, we are not able to distinguish the first and second configurations by using the skyrmion number. Another problem is connected to the fact that if we rely only on $Q$ and the spin-structure factor, the second state may be associated with double-\textit{q} skyrmion state, which is not the case. The developed network approach allows to overcome this classification problem.

Figure~\ref{str2} demonstrates the examples of the non-periodic skyrmion phase at the low temperature and high temperature pure skyrmion and mixed skyrmion-bimeron phases. As can be seen, all of them have the same smeared spin structure factors and approximately equal skyrmion numbers. This makes it impossible to distinguish them by using common techniques. At the same time the developed neural network provides an excellent classification without requiring significant time costs for the calculations.

\section{Machine learning details}

As an input of our FFN we used the $z$ components of the spins obtained from Monte Carlo simulations, then the input and output of the hidden layer neurons were calculated by the following equations

\begin{equation}\label{hi}
h^{inp}_j= \frac{1}{\sum_{i=1}^{N}S_i^z}\cdot\sum_{i=1}^{N}S_i^z W^h_{ij},
\end{equation}
\begin{equation}\label{ho}
h^{out}_j=sigmoid(h^{inp}_j)=\frac{1}{1+e^{-h^{inp}_j}},
\end{equation}
where $S_i^z$ is the value of $i$th input neuron, $W^h_{ij}$ --- weight between the $i$th input neuron and $j$th hidden neuron, $N=L\times L$ --- number of the input neurons. The normalization factor in the first equation is required in order to shift the input value into the range where $sigmoid(h^{inp}_k)\in[0; 1]$. It is very important especially at the beginning of the learning process when we randomly initialize all the weights in range $[-1; 1]$. Without normalization we will obtain  $h^{out}_j$ are equal to 1 or 0 because of the large number of the input units. It will lead to the situation when weights between the hidden and output neurons become the only parameters that affect the result. The values of the output layer neurons were calculated in a standard way by using the following equation

\begin{equation}\label{o}
o_k=sigmoid\left(\sum_{j=1}^{N_h}h^{out}_j W^o_{jk}\right),
\end{equation}
where $N_h$ is the number of hidden neurons, $W^o_{jk}$ --- weight between the $j$th hidden neuron and $k$th output neuron.

During the learning process, we randomly chose 10\% of training set for cross-validation to avoid overfitting and define the stopping point where error is less than the required value. The error function is given by
\begin{equation}\label{Error}
E(o^{ideal},o^{actual})= \frac{\sum_{k=1}^{N_o}(o_k^{ideal}-o_k^{actual})^2}{N_o},
\end{equation}
where $N_o$ is the number of the output neurons, $o^{ideal}$ represents the training labels and $o^{actual}$ is the calculated values of the output neurons.

\begin{figure}[b] 
	\center 
	\includegraphics[width=\columnwidth]{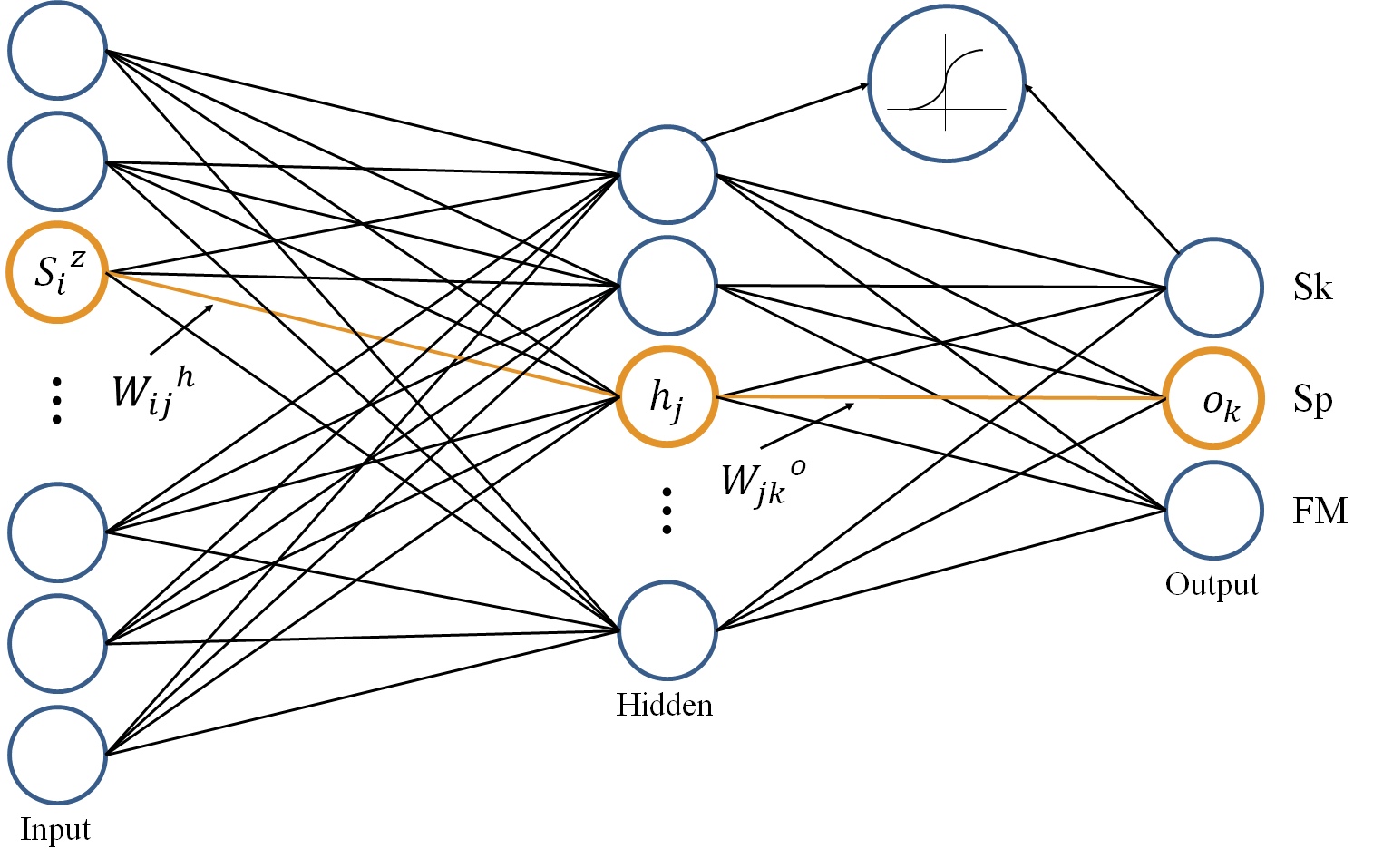}
	\caption{Schematic representation of constructed neural network with single hidden layer. We used sigmoid as an activation function of hidden and output neurons. All the notations are described in the text.}
	\label{netw} 
\end{figure}

Due to the fact that we optimized our network through back-propagation method \cite{backprop} by means of the stochastic gradient descent with momentum we used the following expression for new weights in order to not get stuck in a local minima 
\begin{equation}\label{wi}
W^{(l)}= W^{(l-1)}+\Delta W^{(l)},
\end{equation}
\begin{equation}\label{dwi}
\Delta W_{jk}^{o(l)}=  \alpha\delta o_k h^{out}_j+\mu\Delta W_{jk}^{o(l-1)},
\end{equation}
\begin{equation}\label{dwi}
\Delta W_{ij}^{h(l)}=  \alpha\delta h^{out}_j S^z_i+\mu\Delta W_{ij}^{h(l-1)},
\end{equation}
where $\mu$ is the momentum and $\alpha$ is the learning rate. These parameters can be chosen by trial and error (in our work we used $\mu=0.3$, $\alpha=0.8$). $\delta o_k$ and $\delta h^{out}_j$ are given by 
 
\begin{equation}\label{do}
\delta o_k= (o_k^{ideal}-o_k)o_k(1-o_k),
\end{equation}
\begin{equation}\label{dh}
\delta h^{out}_j= h^{out}_j(1-h^{out}_j)\sum_{k=1}^{N_o}W^o_{jk}\delta o_k.
\end{equation}


\begin{thebibliography}{0}
\bibitem{Troyer}
G. Carleo and M. Troyer,
Science \textbf{355}, 602 (2017).

\bibitem{Imada}
Y. Nomura, A. S. Darmawan, Y. Yamaji, M. Imada,
Phys. Rev. B \textbf{96}, 205152 (2017).

\bibitem{Kato}
H. Saito, M. Kato,
J. Phys. Soc. Jpn. \textbf{87}, 014001 (2018).

\bibitem{phase1}
P. Broecker, J. Carrasquilla, R. G. Melko, and S. Trebst, Sci.
Rep. \textbf{7}, 8823 (2017).

\bibitem{phase2}
K. Ch\textsc{\char13}ng, J. Carrasquilla, R. G. Melko, and E. Khatami,
Phys. Rev. X \textbf{7}, 031038 (2017).

\bibitem{phase3}
L. Wang, Phys. Rev. B \textbf{94}, 195105 (2016).

\bibitem{phase4}
Y. Zhang, R. G. Melko, and E.-A. Kim, Phys. Rev. B \textbf{96},
245119 (2017).

\bibitem{phase5}
E. P. L. van Nieuwenburg, Y.-H. Liu, and S. D. Huber, Nat.
Phys. \textbf{13}, 435 (2017).

\bibitem{phase6}
Y. Zhang and E.-A. Kim, Phys. Rev. Lett. \textbf{118}, 216401
(2017).

\bibitem{phase7}
K. Ch\textsc{\char13}ng, N. Vazquez, and E. Khatami,
Phys. Rev. E \textbf{97}, 013306 (2018).

\bibitem{Melko1}
J. Carrasquilla, R. G. Melko,
Nature Physics \textbf{13}, 431 (2017).

\bibitem{Melko2}
M. J. S. Beach, A. Golubeva, R. G. Melko, 
Phys. Rev. B \textbf{97}, 045207 (2018).

\bibitem{Wessel}
P. Suchsland, S. Wessel, 
Phys. Rev. B \textbf{97}, 174435 (2018).

\bibitem{image}
Y. LeCun, Y. Bengio, G. Hinton, Nature \textbf{521}, 436 (2015).

\bibitem{Bogdanov1}
A. Bogdanov, A. Hubert, Phys. Stat. Sol. (b) \textbf{186}, 527 (1994).

\bibitem{Bogdanov2}
A. Bogdanov, A. Hubert, J. Magn. Magn. Mater. \textbf{138}, 255 (1994).

\bibitem{skyrmion1}
S. M\"uhlbauer, B. Binz, F. Jonietz, C. Pfleiderer, A. Rosch, A. Neubauer, R. Georgii, P. B\"oni, Science \textbf{323}, 915 (2009).

\bibitem{skyrmion2}
F. Jonietz, S. M\"uhlbauer, C. Pfleiderer, A. Neubauer, W. M\"unzer, A. Bauer, T. Adams, R. Georgii, P. B\"oni, R. A. Duine, K. Everschor, M. Garst, A. Rosch,
Science \textbf{330}, 1648 (2010).

\bibitem{skyrmion3}
A. Neubauer, C. Pfleiderer, B. Binz, A. Rosch, R. Ritz, P. G. Niklowitz, and P. B\"oni,
Phys. Rev. Lett. \textbf{102}, 186602 (2009).

\bibitem{expdisk}
F. Zheng, H. Li, S. Wang, D. Song, C. Jin, W. Wei, A. Kov\'acs, J. Zang, M. Tian, Y. Zhang, H. Du, R. E. Dunin-Borkowski, Phys. Rev. Lett. \textbf{119}, 197205 (2017).

\bibitem{FeGe1}
X. Z. Yu, N. Kanazawa, Y. Onose, K. Kimoto, W. Z. Zhang, S. Ishiwata, Y. Matsui, Y. Tokura,
Nature Materials \textbf{10}, 106 (2011).

\bibitem{FeGe2}
M. Nagao, Y.-G. So, H. Yoshida, K. Yamaura, T. Nagai, T. Hara, A. Yamazaki, and K. Kimoto, Phys. Rev. B \textbf{92}, 140415 (2015).

\bibitem{FeIr}
S. Heinze, K. von Bergmann, M. Menzel, J. Brede, A. Kubetzka, R. Wiesendanger, G. Bihlmayer, S Bl\"ugel, Nature Physics \textbf{7}, 713–718 (2011).

\bibitem{MnGe}
N. Kanazawa, J.-H. Kim, D. S. Inosov, J. S. White, N. Egetenmeyer, J. L. Gavilano, S. Ishiwata, Y. Onose, T. Arima, B. Keimer, and Y. Tokura, Phys. Rev. B \textbf{86}, 134425 (2012).

\bibitem{Yu}
X. Z. Yu, Y. Onose, N. Kanazawa, J. H. Park, J. H. Han, Y. Matsui, N. Nagaosa, Y. Tokura, Nature \textbf{465}, 901–904 (2010).

\bibitem{Danis}
D. I. Badrtdinov, S. A. Nikolaev, M. I. Katsnelson, and V. V. Mazurenko,
Phys. Rev. B \textbf{94}, 224418 (2016).

\bibitem{Berg}
Berg, B., L\"uscher, M. Nuclear Physics B \textbf{190}, 412 (1981).

\bibitem{Stamps}
M. C. Ambrose and R. L. Stamps,
New Journal of Physics \textbf{15}, 053003 (2013).

\bibitem{bimeron1}
S. El Hog, A. Bailly-Reyre, H. T. Diep, J. Magn. Magn. Mater. \textbf{452}, 32-38 (2018).

\bibitem{bimeron2}
I. A. Iakovlev, O. M. Sotnikov, V. V. Mazurenko, Phys. Rev. B \textbf{97}, 184415 (2018).

\bibitem{kNN}
F. Pedregosa, G. Varoquaux, A. Gramfort, V. Michel, 
B. Thirion, O. Grisel, M. Blondel, P. Prettenhofer, 
R. Weiss, V. Dubourg, J. Vanderplas, A. Passos, D.
Cournapeau, M. Brucher, M. Perrot, and E. Duchesnay, Journal of Machine Learning Research \textbf{12}, 2825-2830 (2011).

\bibitem{backprop}
MacLeod C. \textit{An Introduction to Practical Neural Networks and Genetic Algorithms for Engineers and Scientists}, Robert Gordon University, Aberdeen, Scotland, -157 pp. (2001).

\end{thebibliography}
\end{document}